\documentclass[aps,showpacs]{revtex4}

\newcommand{\eqref}[1]{(\ref{#1})}
\renewcommand{\d}{\partial}
\newcommand{\nn}{\nonumber\\}

\newcommand{\exvs}[1]{\langle{#1}\rangle}
\newcommand{\ep}{\varepsilon}

\renewcommand{\k}{{\bf k}}
\newcommand{\p}{{\bf p}}

\newcommand{\rh}{\varrho}
\newcommand{\T}{\textrm{T}}

\newcommand{\Disc}{\mathop{\textrm{Disc}}}
\renewcommand{\Im}{\,\textrm{Im}\,}
\renewcommand{\Re}{\,\textrm{Re}\,}
\newcommand{\pint}[2]{{\int\!\frac{d^{#1}#2}{(2\pi)^{#1}}\,}}
\newcommand{\MSbar}{{\ensuremath{\overline{\mathrm{MS}}}}}

\newcommand{\sgn}{{\ensuremath{\mathop{\mathrm{sgn}}}}}

\usepackage{graphicx}

\begin{document}

\title{Renormalization in momentum dependent resummations}
\author{Antal Jakov\'ac} 
\email{Antal.Jakovac@cern.ch}
\affiliation{Institute of Physics, Budapest University of Technology
  and Economics, Budafoki \'ut 8, H-1111 Budapest , Hungary}
\date{\today}

\begin{abstract}
  At finite temperature and in non-equilibrium environments we have to
  resum perturbation theory to avoid infrared divergences. Since
  resummation shuffles the perturbative orders, renormalizability is a
  nontrivial issue. In this paper we demonstrate that one can modify
  any type of resummations -- even if it is momentum dependent, or it
  resums higher point functions -- in a way that it is renormalizable
  with the usual zero temperature counterterms.  To achieve this goal
  we reformulate resummation in form of a renormalization scheme. We
  apply this technique to perform a 2PI resummation in the six
  dimensional cubic scalar model in equilibrium.
\end{abstract}


\maketitle

\section{Introduction}
\label{sec:intro}

Renormalized perturbation theory in \MSbar\ scheme was developed to
extract perturbative information about vacuum expectation values of
observables at high energies. Still, when it fails, for example in
computation of expectation values with a generic density matrix, we
tend to speak about ``non-perturbative effects'' and the failure of
perturbation theory (PT). An example at high temperature when this
pathological behavior occurs was demonstrated by Dolan and Jackiw
\cite{DolanJackiw}, showing that in $\Phi^4$ theory at finite
temperatures higher order diagrams can provide contributions
$\sim(\lambda T^2/m^2)^n$ which is large at high temperatures. They
also found the solution of this problem: with (re)summation of a
specific set of diagrams (the daisy diagrams), the problematic
$\sim\lambda T^2$ term becomes part of the propagator mass, thus
appearing in the denominator instead of the numerator.

This solution reveals, at the same time, the core of the problem.
Perturbation theory in \MSbar\ scheme starts with a free action,
fitted to the zero temperature particle spectrum. At finite
temperatures the true excitation spectrum changes significantly. Since
particle spectrum comes from the free part of the Lagrangian, any
perturbation theory will fail which tries to follow this change via
interactions. This line of thought also provides an easier way of
performing daisy resummation, by changing particle spectrum of the
theory from the beginning. Technically one adds to and subtracts from
the Lagrangian the same (temperature dependent) mass term, treating
one as a part of the free action, the other as a (thermal) counterterm
\cite{thermalcount}. The modification mass term can be set equal to
the high temperature limit of the self energy diagram, or it can be
optimized via a gap equation \cite{Buchmuller_gapeq, Patkos_optpert}.

Taking into account thermal effects only through an effective mass can
be used only cautiously in case of more complicated situations, as in
non-equilibrium, or in gauge theories\footnote{In gauge theories the
  introduction of magnetic mass is an example where the effective mass
  can have an important physical role}. In these cases the
non-trivial, non-quadratic momentum dependence of the self energy is
important, too, and it must be taken into account already in the free
Lagrangian. In gauge theories the hard thermal loop (HTL) improved
perturbation theory \cite{BraatenPisarski} provides this type of
resummation, which uses the high temperature limit of the momentum
dependent self energy as part of the effective propagator (and uses
also effective vertices to maintain gauge invariance, cf.
\cite{BraatenPisarski}). In the generic case the modification of the
spectrum can be taken into account most effectively with the method of
Cornwall, Jackiw and Tomboulis \cite{CJT}, which performs
two-particle-irreducible (2PI) resummation.  In this method we should
compute the diagrams with the exact (also background field dependent)
propagator, and, consequently, no self-energy corrections should be
taken into account. This method was intensively developed and used in
numerical calculations in the recent literature
\cite{CJT_developement}.

The consequent application of all these methods is obstructed by their
non-trivial relation to the renormalization. Direct application of the
thermal counterterm method, for example, requires temperature
dependent infinite counterterms to cancel the one-loop level $\sim
m_T^2/\ep$ type of divergences, where $m_T^2$ is the thermal mass
\cite{Patkos_optpert, Caldas, JakovacSzep}. The temperature dependent
piece drops out at two loop level, but there new temperature dependent
divergences are generated. The situation is even more serious in the
case of 2PI method: at one loop level we generate terms to the exact
propagator that are $\sim m^2 \log p$; when putting back into loops,
this term provides us with new types of divergences, not present in
the original renormalized Lagrangian. Therefore the resummation
obtained by truncating the 2PI equations at a given loop order is not
consistent as a renormalizable theory.

The main reason of failure is that we separate the loop order of the
IR sensitive diagram we want to resum and the corresponding
counterterm diagrams, which, in the original PT, made them finite. In
a resummation based on fixed order skeleton diagrams this leads to
uncompensated divergences. There are different methods to keep track
of the divergences. The local case is simpler and there we have a
deeper understanding of the problem. A self-consistent method for
resummation was developed in \cite{BanerjeeMallik,Caldas} to obtain
finite gap equations. In \cite{Verschelde} the self-consistent
resummation in the framework of 2PPI effective action is accomplished
by explicitely introduced counterterms into the Lagrangian. In
\cite{JakovacSzep} this method is generalized to any,
momentum-independent resummations. Here an alternative viewpoint is
applied: the resummation being treated as a change of renormalization
scheme. The idea of adapting the renormalization process to the
environment was first suggested in \cite{envfriend}.

The momentum dependent resummations are more difficult, here the
results are concentrated on the 2PI resummation. In a seminal paper,
van Hees and Knoll \cite{HeesKnoll} presented a proof based on the
BPHZ scheme showing that it is in fact enough to use the zero
temperature BPHZ scheme to renormalize the finite temperature 2PI
resummation. The central problem there is to renormalize the zero
temperature ladder resummation, ie. the Bethe-Salpeter equations. This
method was further developed and used in \cite{Blaizotetal, other2PI,
  Bergesetal}. A method to renormalize Hartree-Fock approximation was
developed in \cite{Destrietal}.

It is, however, still an open question, what kind of momentum
dependent resummations are renormalizable. In this paper we study this
question, primarily in the framework of the four dimensional $\Phi^4$
and six dimensional $\Phi^3$ model, but the logic can be easily
extended to more complicated models. The starting point is the method
of \cite{JakovacSzep}, which is generalized to momentum dependent
case. The main idea is that, instead of resumming the necessary
counterterm diagrams in the original resummation, we slightly change
the resummation procedure: in the low energy (IR) regime it will be
still the desired resummation, while at high energies (UV) it falls
back to the ordinary perturbation theory. This means that the
divergences remain identical to the usual zero temperature ones.

The paper is structured as follows. In Section \ref{sec:reinterpret}
we give a different view of resummation, interpreting it as a specific
renormalization scheme even in the momentum dependent case. We
generalize somewhat the notion of ``renormalized Lagrangian'',
allowing momentum dependent parameters, and taking care not to spoil
the perturbative renormalizability. In Section \ref{sec:ren2PI} we
summarize the results of 2PI resummation, and find that specific
resummation scheme that corresponds to it. It will be shown that it is
not a well-defined scheme (as it can be conjectured from the
renormalization problems of the 2PI resummation itself), so in Section
\ref{sec:improved} we suggest improvements of the method that is
capable to satisfy renormalizability and 2PI resummation in the IR
regime at the same time. In Section \ref{sec:Phi3} we apply the method
to find the 2PI resummed self energy of the six dimensional $\Phi^3$
model. In Section \ref{sec:Conclusions} we give our concluding remarks
and discuss open questions.

\section{Reinterpreting resummation}
\label{sec:reinterpret}

Resummation is a notion that pre-assumes a singled out perturbation
theory (the ``unresummed'' one). It suggests that in this singled out
perturbation theory (PT) we have to do additional work to compute some
observables by taking into account contributions from higher orders.
Usually this singled out PT is the one stemming from the \MSbar\ 
renormalized Lagrangian. This works nicely when used for calculation
of vacuum expectation values of high energy observables, but needs to
be resummed, for example, at finite temperatures.

This way of thinking is, however, misleading, as we have no a priori
reason to prefer one environment/process above others. Having in
mind this democracy of environments, we should not speak about
resummation, but about perturbation theories best adapted to a certain
environment (cf. also \cite{envfriend}).

The definition of quantum field theory is based on the path integral
living on a fine enough spacetime lattice, with an action (the bare
action) that have parameters depending on the lattice spacing. This UV
action provides, in principle, the expectation values of any IR
observables, at any environment. If we do PT, however, we must have
some knowledge about the result in the specific environment in order
that already the first approximation is ``sufficiently'' close to the
exact result: ie. we should ensure that the corrections are indeed
``perturbative'', smaller than the leading order. Since in the usual
PT the solvable part is the free (quadratic) theory, the above
requirement says that the vacuum as well as the free excitation
spectrum should be sufficiently close to the exact one. But the vacuum
of the theory may change in different phases of the matter, and also
the excitation spectrum can be dramatically different at different
physical environments, for example at different temperatures, not to
speak about the non-equilibrium environment. So, although it is
implicit in the usual discussions, before we start PT, half of the
task should be solved: to guess the vacuum and the spectrum
sufficiently well.

Different PTs, adapted to different environments, yield different
results already at tree level. But, of course, at infinite order any
consistent PT should provide the same result for a given observable.
That means that the result of the tree level of PT $A$ is somehow
hidden in the higher orders of PT $B$. From the point of view of $B$,
therefore, $A$ represents a resummation of the perturbative series --
and vice versa.  And both $A$ and $B$ are resummations of the bare
perturbative series. A zero temperature example is that results
obtained in the on-mass shell scheme seem to be ``resummed'' from the
point of view of the \MSbar\ scheme.

In addition to the necessity of resummation there is another problem.
When we start with a UV theory with appropriate bare parameters, it is
granted that any IR observables are ``finite'', ie. do not depend on
the exact value of the UV cutoff. The requirement for a consistent PT
is, however, much more severe: we should have finite results order by
order. Only a very smart resummation of the bare theory, known as the
renormalized PT, can provide finite results at each perturbative order
\cite{Collins}.  No other perturbative methods are known that would be
capable to render the theory IR finite. Therefore, if
renormalizability is an issue, we must not use other PTs than the
renormalized ones.

Technically, in renormalized PT, we split the parameters into a
renormalized and counterterm part, the latter being cutoff dependent,
and should formally be taken into account at higher orders of the
calculation. Renormalizability requires specific form only for the UV
part. Thus it allows to work with different PTs, which use different
finite parts of the counterterms: these are called renormalization
schemes (RSs). When we want to fit the PT to the external environment
the only thing we can play with is the finite part of the
counterterms; ie. we should adapt the RS to the environment by
choosing appropriate finite parts.

\subsection{Class of constant physics}
\label{sec:clconstphys}

In the family of RSs we can define an equivalence relation: two RSs
are equivalent if they provide the same expectation values for any
$n$-point functions at infinite order (up to a finite normalization
factor). This leads to the equivalence classes of schemes that
provides ``constant physics''.

Let us take two RSs: $A$ with parameters $g^A_i$, and $B$ with
parameters $g^B_i$. The expectation value of an $n$-point
function $\hat O_n= \T \Phi(x_1)\dots \Phi(x_n)$ will depend on these
parameters:
\begin{equation}
  \exvs{\hat O_n}^A = O^A_n(g^A),\qquad  \exvs{\hat O_n}^B =
  O^B_n(g^B).
\end{equation}
$A$ and $B$ belong to the same class of constant physics, if we have
\begin{equation}
  O^A_n(g^A) = \zeta_{AB}^n O^B_n(g^B),
\end{equation}
with a ($n$-independent) constant $\zeta_{AB}$. According to the theory of
renormalization \cite{Collins} these equations can be fulfilled
\emph{for any $n$}, if there is a specific relation between the
renormalized parameters
\begin{equation}
  \label{RGgeneral}
  g^A_i = f^{AB}_i(g^B).
\end{equation}
Sometimes these equations are referred to as the renormalization group
(RG) equations. The physical condition for providing the same values
for the observables from both schemes is that they both stem from the
same bare Lagrangian \cite{Collins}.

To demonstrate these notions we give an example in the $\Phi^4$
model. The Lagrangian reads in terms of the bare quantities (denoted
by subscript ``0''):
\begin{equation}
  \label{barePhi4}
  {\cal L} = \frac12 (\d\Phi_0)^2 - \frac {m_0^2}2\Phi_0^2
  -\frac{\lambda_0}{24} \Phi_0^4.
\end{equation}
The renormalized Lagrangians are used to write in the following form
\begin{equation}
  \label{renPhi4}
  {\cal L} = \frac12 (\d\Phi)^2 - \frac {m^2}2\Phi^2
  -\frac{\lambda}{24} \Phi^4 + \frac{\delta Z^2} 2 (\d\Phi)^2 - \frac
  {\delta m^2}2\Phi^2 -\frac{\delta \lambda}{24} \Phi^4,
\end{equation}
where the formal power of $\lambda$ in the counterterms determine the
loop order they first contribute. Since \eqref{barePhi4} determines,
nonperturbatively, the expectation values of all IR observables,
clearly the renormalized Lagrangians of \eqref{renPhi4} are giving the
same value (at infinite order), if we satisfy
\begin{equation}
  \Phi_0 = Z \Phi, \qquad Z^2m_0^2 = m^2+\delta m^2,\qquad
  Z^4\lambda_0= \lambda+\delta\lambda,
\end{equation}
where $Z^2=1+\delta Z^2$. Therefore two schemes give the same result (up
to normalization), if they obey:
\begin{equation}
  Z_A = \zeta_{AB} Z_B,\qquad  m_A^2+\delta m_A^2 = \zeta_{AB}^2( m_B^2+\delta
  m_B^2), \qquad \lambda_A+\delta\lambda_A =
  \zeta_{AB}^4(\lambda_B+\delta\lambda_B).
\end{equation}

The condition that schemes $A$ and $B$ belong to the same class of
constant physics can be rewritten compactly, using the generator of 1PI
diagrams as
\begin{equation}
  \label{RGGamma}
  \Gamma_A[\Phi] = \Gamma_B[\zeta_{AB} \Phi].
\end{equation}

At finite $n$th order this is satisfied up to order $g^n$. Then the
left hand side contains contributions from higher orders, as
considered from RS $B$. So the left hand side provides a resummation
for RS $B$. Moreover, since all the functions $f^{AB}$ and $O^A$ are
finite, we have a finite, ie. renormalized resummation.

\subsection{Momentum independent resummations}
\label{sec:momindres}

Let us try to apply the ideas of the previous subsection to the
momentum independent mass resummation in $\Phi^4$ model. A more
comprehensive study of this case can be found in Ref.
\cite{JakovacSzep}.

Let us choose for RS $A$ the \MSbar\ scheme. At one loop level with
dimensional regularization the counterterms are defined as
\cite{JakovacSzep}:
\begin{equation}
  \delta Z_{\MSbar,1}=0,\qquad  \delta m_{\MSbar,1}^2 = -m^2
    \frac\lambda{32\pi^2} \left[-\frac1\ep +\gamma_E-1 - \ln 4\pi
    \right],\qquad 
    \delta\lambda_{\MSbar,1} =
    -\frac{3\lambda^2}{32\pi^2}\left(-\frac1\ep +\gamma_E
    -\ln4\pi\right).
\end{equation}
In this scheme zero temperature observables can be calculated with
reasonable precision; at finite temperatures, however, IR divergences
make the calculation difficult. For example the 1-loop self energy
at zero momentum reads, in dimensional regularization:
\begin{equation}
  \Pi_\MSbar(p=0) = m^2 + \frac{\lambda m^2}{32\pi^2} \left[-\frac1\ep
    +\gamma_E-1 +\ln\frac{m^2}{4\pi\mu^2} \right] + \frac\lambda 2
    J(m) + \delta m_{\MSbar,1}^2,
\end{equation}
where $\mu$ is the renormalization scale. $J$ is the finite temperature
part of the bosonic tadpole function:
\begin{equation}
  \label{Jm}
  J(m) = \frac1{2\pi^2}\int\limits_{m}^\infty\!
    d\omega\,\sqrt{\omega^2-m^2}\,n(\omega),
\end{equation}
where $n(\omega)$ is the Bose-Einstein distribution. It can be
approximated at high temperatures as
\begin{equation}
  J(m) \approx \frac{T^2}{12} -\frac {Tm}{4\pi} - \frac {m^2}{16\pi^2}
  \ln\frac{m^2}{cT^2}+ \dots, 
\end{equation}
where $\ln c=1+2\ln4\pi-2\gamma_E\approx4.90762$.  Therefore, at high
temperatures, the one-loop self energy will be
\begin{equation}
  \Pi_\MSbar(p=0,T) = m^2 +  \frac{\lambda T^2}{24} - \frac{\lambda T
  m}{8\pi} + \frac{\lambda m^2}{32\pi^2} \ln \frac{cT^2}{4\pi\mu^2}.
\end{equation}
This means that the first order correction can be larger than the
tree level term, if $\lambda T^2\gg m^2$, ie. at high temperatures. At
higher orders this failure of the \MSbar\ scheme becomes more and more
obvious, as we receive $(\lambda T^2/m^2)^n$ contributions from the
daisy diagrams \cite{DolanJackiw}.

This phenomenon, however, does not mean the failure of perturbation
theory itself, since we can easily find a RS, where this problem does
not show up. Indeed, the finite temperature version of the
on-mass-shell (M) scheme uses the counterterm
\begin{equation}
  \label{thermalmasscount}
  \delta m_{M1}^2 = -\frac{\lambda m^2}{32\pi^2} \left[-\frac1\ep
    +\gamma_E-1 +\ln\frac{m^2}{4\pi\mu^2} \right] -\frac\lambda2 J(m),
\end{equation}
and so the self energy in this scheme reads simply
\begin{equation}
  \Pi_M(p=0,T) = m^2.
\end{equation}
This is just the tree level value -- as we expect from the
on-mass-shell scheme. No dangerous $\lambda T^2$ terms appear, and by
definition it extends to all orders in PT. So it seems to be much
better to use the on-mass-shell scheme at high temperatures for
perturbative calculations.

If we want to compare the results of the \MSbar\ and the finite
temperature on-mass-shell schemes -- for example if we want to follow
the temperature dependence of the quantities starting from zero
temperature -- then we must take care that they belong to the same
class of constant physics. Since now only the mass counterterm is
changed, we have $\zeta=1$ and  $\lambda_\MSbar=\lambda_M =\lambda$,
and we must satisfy
\begin{equation}
  m_\MSbar^2 - \frac{\lambda m_\MSbar^2}{32\pi^2} \left[-\frac1\ep
    +\gamma_E-1 - \ln 4\pi \right] + {\cal O}(\lambda^2)= m_M^2 -
    \frac{\lambda m_M^2}{32\pi^2} \left[-\frac1\ep +\gamma_E-1
    +\ln\frac{m_M^2}{4\pi\mu^2} \right] - \frac\lambda2 J(m_M) + {\cal
    O} (\lambda^2). 
\end{equation}
At ${\cal O}(\lambda)$ this simply yields
\begin{equation}
  \label{eq:tadpolegapeq}
  m_\MSbar^2 = m_M^2 - \frac{\lambda m_M^2}{32\pi^2}
    \ln\frac{m_M^2}{\mu^2} -\frac\lambda2 J(m_M).
\end{equation}
When we solve this equation -- a finite gap equation -- the result
will be applicable in the given class of constant physics. In
particular, we can interpret this result as a resummed \MSbar\ self
energy:
\begin{equation}
  \label{eq:tadpoleresum}
  \Pi_\MSbar^{resummed}(p=0,T) = m_M^2(m_\MSbar,\lambda).
\end{equation}

An alternative to this approach is when we define a scheme where only
the dangerous finite temperature part is resummed. This $F$ scheme is
defined by the counterterm
\begin{equation}
  \label{thermalmasscountF}
  \delta m_{F1}^2 = -\frac{\lambda m^2}{32\pi^2} \left[-\frac1\ep
    +\gamma_E-1 - \ln4\pi \right] -\frac\lambda2 J(m),
\end{equation}
and so the self energy in this scheme reads simply
\begin{equation}
  \Pi_F(p=0,T) = m^2\left[ 1 + \frac{\lambda}{32\pi^2} \ln
  \frac{m^2}{\mu^2} \right].
\end{equation}
This is the zero temperature \MSbar\ value, and we avoided the
dangerous $\lambda T^2$ terms again. In this case the condition of the
constant physics reads:
\begin{equation}
  m_\MSbar^2 - \frac{\lambda m_\MSbar^2}{32\pi^2} \left[-\frac1\ep
    +\gamma_E-1 - \ln 4\pi \right] + {\cal O}(\lambda^2)= m_F^2 -
    \frac{\lambda m_F^2}{32\pi^2} \left[-\frac1\ep +\gamma_E-1
    -\ln4\pi \right] - \frac\lambda2 J(m_F) + {\cal O}(\lambda^2),
\end{equation}
which, at ${\cal O}(\lambda)$, implies
\begin{equation}
  \label{eq:tadpolegapeqF}
  m_\MSbar^2 = m_F^2 -\frac\lambda2 J(m_F).
\end{equation}
This is another, also usual, form of the gap equation. The resummation
again reads as
\begin{equation}
  \label{eq:tadpoleresumF}
  \Pi_\MSbar^{resummed}(p=0,T) =
  \Pi_F(p=0,T)\biggr|_{m_F(m_\MSbar,\lambda)}.
\end{equation}
This is a finite version of the thermal mass method.

\subsection{Momentum dependent schemes}
\label{sec:momdep}

What was said above is perfectly applicable to momentum independent
resummations, as it was demonstrated in Ref. \cite{JakovacSzep}. To be
able to treat momentum dependence we have to rethink what are really
the conditions for a Lagrangian to be renormalizable.

In fact, all theorems of renormalizability are based on the behavior
of the Feynman diagrams when internal momenta are asymptotically
large. One should prove that a diagram can be divergent if some of its
internal momenta go to \emph{infinity} on the same rate
\cite{Collins}; we do not have new divergences when different momenta
go to infinity at different rate. One proves that after removing the
divergences of all sub-diagrams the diagram is superficially divergent
with polynomial momentum dependence. Then the BPHZ scheme and forest
formula go through, and we can perturbatively renormalize the theory
\cite{Collins}.

It is clear that all these theorems still hold, if propagators or
vertices are momentum dependent, but in a way that the
\emph{asymptotic} behavior is the same as in the usual case. More
precisely, the new momentum dependence should not cause any new
divergences in any diagrams.

For example, if we want to use momentum dependent ``mass'' term in
$\Phi^4$ theory, we have to take into account, that the most singular
diagram contains one propagator and one integration (the
tadpole). If the mass has asymptotic momentum dependence as
\begin{equation}
  \label{m2p}
  m^2(p) = m_R^2 + {\cal O}\left(p^{-\gamma}\right),
\end{equation}
then the divergence of the tadpole diagram calculated with this
``mass'' reads
\begin{equation}
  \pint4p \frac1{p^2-m^2(p)} = \pint4p \frac1{p^2-m_R^2
  -{\cal O}\left(p^{-\gamma}\right)} \longrightarrow \pint4p\left[
  \frac1{p^2-m_R^2} + {\cal O}\left(\frac{p^{-\gamma}} {(p^2-m_R^2)^2}
  \right)\right].
\end{equation}
Here the second term yield finite result if $\gamma>0$; then it is not
relevant for the issue of renormalizability. As a consequence we are
allowed to use any momentum dependent mass term that approaches
constant value polynomially at asymptotic large momenta.

This means that we can define a consistent renormalization scheme,
by splitting the bare mass term as
\begin{equation}
  Z m_0^2 =  m^2(p) + \delta m^2(p)
\end{equation}
where $m^2(p)$ satisfies \eqref{m2p}.

If the coefficient of the quadratic term is considered, ie. $p^2-m^2$,
then the allowed modification is ${\cal O}(p^{-2-\gamma})$ relative to the
leading $p^2$ term. A similar observation can be made for coefficients
of higher powers. For example, if the quadratic coupling is denoted as
$\lambda$, then we can introduce a momentum dependent coupling as
\begin{equation}
   \frac1{24}\,Z^2{\lambda_{bare}}\, (2\pi)^4\delta(\Sigma p_i)
   \,\Phi(p_1) \Phi(p_2) \Phi(p_3) \Phi(p_4) = \frac1{24}\left(
   \lambda(p_i) + \delta \lambda(p_i)\right)\, (2\pi)^4\delta(\Sigma
   p_i) \,\Phi(p_1) \Phi(p_2) \Phi(p_3) \Phi(p_4).
\end{equation}
If at high momenta the asymptotic behavior of $\lambda_R$ is denoted
as
\begin{equation}
  \lambda(p_1,p_2,p_3,p_4) = \lambda_R + {\cal O}((\max p_i)^{-2-\gamma}),
\end{equation}
then the tadpole diagram is computed as 
\begin{equation}
  \pint4p \frac{\lambda(p,q,k,\ell)} {p^2-m^2(p)} \longrightarrow
  \pint4p\left[ \frac{\lambda_R} {p^2-m_R^2} + {\cal O}
  \left(\frac{p^{-2-\gamma}} {p^2-m_R^2} \right)\right],
\end{equation}
where the second term is convergent if $\gamma>0$.

So as a general conclusion we can state that a Lagrangian remains
perturbatively renormalizable if the couplings are momentum dependent,
provided they are suppressed by a factor ${\cal O}(p^{-2-\gamma})$
relative to the leading momentum power at high momenta. 

Finally we give two remarks. First, the order ${\cal O}(p^{-2-\gamma})$
claims nothing about logarithmic correction, since it has sub-power
asymptotics. So, for example, an asymptotic behavior $\log(p)\,
p^{-2-\gamma}$ is still allowed. The second remark is that we may not
stop at the level of the usual renormalizable operators. We can add
$\sim \Phi^6$ terms with coefficient of the leading term ${\cal O}
(p^{-2-\gamma})$. This makes the renormalizable treatment of nPI
diagrams possible.

\subsection{Momentum dependent renormalizable resummations}
\label{sec:momdepres}

Now we are ready to generally describe the resummations that are
renormalizable at the same time. We restrict ourselves here to
self-energy type resummations.

We first define a scheme (M) that is particularly adequate for the
problem at hand. In this case it is simply the choice of the momentum
dependent mass counterterm $\delta m_M^2(p)$. Anticipating matching to
\MSbar\ scheme, we also use momentum dependent mass $m^2(p)$. To
simplify the notation we introduce the renormalized free propagator in
this scheme as
\begin{equation}
  G^{-1}_{M,0}(p) = p^2 - m^2(p).
\end{equation}
Now we perform perturbation theory in this scheme. Since it is a well
defined scheme, all expectation values are finite. In particular the
1PI generating functional $\Gamma_M[\Phi]$ is finite, too. This will
depend on the propagator; to emphasize this dependence we shall write
$\Gamma_M[\Phi;G_{M,0}]$.

Now we translate the results to \MSbar\ scheme. Since only the masses
are modified (that is $\delta Z^2$ and $\delta \lambda$ are the same as 
in \MSbar) , the condition of the constant physics reads simply:
\begin{equation}
  m_\MSbar^2 + \delta m_\MSbar^2 = m_M^2(p) + \delta m_M^2(p).
\end{equation}
In practical cases $\delta m_M^2(p)$ depends on the propagator
$G_{M,0}(p)$, and so this is a gap equation. Since the left hand side
is momentum independent, $m_M^2(p)$ should compensate the momentum
dependence of $\delta m_M^2(p)$. If the counterterm has a polynomially
vanishing momentum dependence, the mass will also have polynomially
vanishing momentum dependence. In this case, according to the analysis
of the previous subsection, the divergences in $\delta m_\MSbar^2$ and
$\delta m_M^2(p)$ are the same.  That means, that we have a finite
relation between $ m_M^2(p)$ and $m^2_\MSbar$ -- or, expressing in
another way, the propagator in scheme $M$ depends explicitly on the
\MSbar\ mass value: $G_{M,0}(p;m_\MSbar)$.

With this dependence all expectation values are the same in \MSbar\
and in $M$ schemes, if we calculated them at infinite order. At finite
$n$th order they are the same only up to $\lambda^n$, the rest is the
``resummation'' as considered from the point of view of \MSbar\
scheme:
\begin{equation}
  \Gamma_\MSbar^{resummed}[\Phi] = \Gamma_M[\Phi;G_{M,0}]
  \biggr|_{G_{M,0}\ \mathrm{from\ constant\ physics}}.
\end{equation}

To make this, in essence very simple, strategy more tractable, we now
work out some examples in the next sections.

\section{Renormalization of 2PI resummation}
\label{sec:ren2PI}

A popular resummation method is the 2PI resummation; for a
comprehensive description and further references cf. \cite{CJT,
  CJT_developement}. The physical idea is that the free theory should
characterize the quasiparticle propagation in the best possible way,
so we should use the exact propagator already in the free theory.
Therefore in perturbation theory the internal lines are exact
propagators, which also implies that it must not contain self-energy
insertions. Since self-energy insertions are separated from the rest
of the diagram by two lines, the above idea is based on diagrams that
are two-particle irreducible. In the usual formalism it is achieved by
introducing a generic propagator as an external quadratic current, and
then perform a Legendre transformation on that. The requirement that
the free propagator is exact can be formulated as a gap equation.

There are different methods to renormalize 2PI resummation; these
usually use BPHZ scheme, and rely on the renormalization of the zero
temperature Bethe-Salpeter kernel \cite{HeesKnoll}. The method
proposed by this paper makes the 2PI effective action finite in a
different way, directly using the (generalized) counterterms of the
original Lagrangian.

\subsection{The 2PI resummation as a renormalization scheme}

In the skeleton diagrams of the 2PI resummation the self-energy
diagrams are missing. We can define a scheme, where this occurs, by
appropriately choosing the mass counterterm. To understand the idea we
recall the zero temperature on-mass shell scheme: here one cancels
radiative mass corrections on the mass shell with help of the finite
part of the mass counterterm, thus achieving that the tree level mass
is the exact mass in the same time. The exact mass reads as
\begin{equation}
  m^2_{ex} = m^2 + \delta m^2 + \Pi(k)|_{k^2 = m_{ex}^2},
\end{equation}
and we have to ensure that it agrees with the free mass, ie.
$m_{ex}^2=m^2$. That yields the following choice of the mass
counterterm
\begin{equation}
  \label{eq:oms}
  \delta m^2 + \Pi(k)|_{k^2 = m^2} = 0,
\end{equation}
therefore the renormalized self-energy correction is zero on the mass
shell. To fully define the scheme, we should choose the finite parts
of the other counterterms ($\delta\lambda,\,\delta Z$): we choose for
them the \MSbar\ values.

We should generalize this scheme in a way that not only the mass, but
the complete exact propagator is the same as the free propagator. We
use a momentum dependent mass $m(k)$, yielding the free propagator
$G^{-1}=k^2 -m^2(k)$, together with momentum dependent mass
counterterm $\delta m(k)$. From Schwinger-Dyson equation, using
explicitly the mass counterterm, we can write
\begin{equation}
  G^{-1}_{ex}(k) = G^{-1}(k) -   G^{-1}_{ex}(k) (\Pi(k) + \delta
  m^2(k)) G^{-1}(k).
\end{equation}
The condition that the free propagator is exact, $G_{ex}(k) = G(k)$,
can be satisfied \emph{for all $k$} if
\begin{equation}
  \label{2PIcnt}
  \delta m^2(k) + \Pi(k) = 0.  
\end{equation}
This is in direct analogy with the on-mass-shell equation
\eqref{eq:oms}. We will call the scheme defined above as bare 2PI 
scheme.

Two remarks are in order here. First we call the attention to the fact
that $\Pi(k)$ implicitly depends on the propagator $G$ so on the
momentum dependent mass $m(k)$, too. Secondly, if the bare 2PI scheme 
was a
consistent scheme (as it is not, see later), the above condition
should determine the finite part only, the regularized UV divergences
should be the same as in any other schemes.

If we want to have the result in \MSbar\ scheme we must ensure that
the 2PI and the \MSbar\ schemes belong to the same class of constant
physics. According to our previous analysis in Section
\ref{sec:momdep} we should provide the same the bare mass in the two
cases:
\begin{equation}
  \label{RGrel}
  Z^{-2}(m^2(k) + \delta m^2(k)) = Z_\MSbar^{-2}(m^2_\MSbar + \delta
  m^2_\MSbar).
\end{equation}
We can choose the same wave function renormalization counterterm in
the two schemes. Then, using \eqref{2PIcnt}, we can write
\begin{equation}
  m^2(k) - \Pi[m,k] = m^2_\MSbar +\delta m^2_\MSbar,
\end{equation}
where we explicitly written out the mass dependence of the
self-energy. Equivalently we can write
\begin{equation}
  \label{2PIscheme_massrel}
  m^2(k) = m^2_\MSbar +  \Pi_\MSbar[m,k].
\end{equation}
Here $\Pi_\MSbar[m,k] = \Pi[m,k] - \delta m^2_\MSbar$, a finite
quantity (if the bare 2PI scheme was a consistent renormalization 
scheme). If we add a $-k^2$ to both sides we find
\begin{equation}
  G^{-1}(k) = G^{-1}_\MSbar -  \Pi_\MSbar[G,k].  
\end{equation}
This equation is exactly the same as the 2PI gap equation that
determines the self energy, or, equivalently, $G(k)$.

\subsection{Improvement of the 2PI resummation}
\label{sec:improved}

Although the formalism of the 2PI resummation is appealing, it is
inconsistent when using order by order. The self energy, namely, can
yield corrections of type $k^2\ln k,\, m^2\ln k$, which, if considered
to be part of the propagator, leads to new, eventually temperature
dependent, divergences in the calculation that cannot be canceled by
usual counterterms. The related problem in the bare 2PI scheme is that
the condition \eqref{2PIscheme_massrel} requires that $m^2(k)$
contains momentum dependence $\sim k^2\ln k$ or $\sim m^2\ln k$, which
is not the allowed ${\cal O}(k^{-\gamma})$ order. This means that the bare
2PI scheme is, in fact, not a consistent renormalization scheme.

In the present framework we can accommodate to the requirement of
renormalizability, if we employ 2PI resummation in the \emph{IR}
regime, and leave the asymptotic momentum regime untouched. These
schemes are already consistent, and we call them 2PI schemes. There
are several possibilities to achieve this.

The first is that we take into account the momentum dependence only
below a scale $M$; above this scale we stay with the above discussed
thermal mass approximation. That is we write for the mass counterterm,
in high temperature approximation (cf. \eqref{thermalmasscount}):
\begin{equation}
  \delta m^2(k) = \left\{
    \begin{array}[c]{ll}
      -\Pi(k)\qquad& \mathrm{if}\, |k| < M\cr
      \delta m_\MSbar^2 - \frac{\lambda T^2}{24},\qquad& \mathrm{if}\,
      |k| > M\cr 
    \end{array}
  \right.
\end{equation}
Since at asymptotically high momenta we recover Lorentz invariance, we
can use for $|k|$ the Euclidean norm. The condition of the constant
physics \eqref{2PIscheme_massrel} now reads as
\begin{equation}
    m^2(k) = m^2_\MSbar +  \Theta(|k|<M)\Pi_\MSbar[m,k] +
    \Theta(|k|>M) \left[\frac\lambda2J(m^2) + \delta m_\MSbar^2\right].
\end{equation}
This describes a scheme that provides 2PI resummation at low momenta,
and super-daisy resummation at high momenta. Since we expect that the
effects of the environment, and so the spectrum modifications are
restricted to the IR regime, this choice hopefully catches the main
physical points.

We can apply, instead of a sharp cutoff, a smeared cutoff function
$M(k)$ that is $1$ for small momenta and vanishes asymptotically as
$k^{-2-\gamma}$ at least. Then we could write
\begin{equation}
  \delta m^2(k) = -M(k)\Pi(k) - (1-M(k)) \frac\lambda2J(m^2),
\end{equation}
and the condition of the constant physics reads
\begin{equation}
    m^2(k) = m^2_\MSbar +  M(k) \Pi_\MSbar[m,k] +
    (1-M(k))\left[\frac\lambda2 J(m^2)+\delta m_\MSbar^2\right].
\end{equation}

Another possibility, related to the technique applied by
\cite{Blaizotetal}, is to subtract the problematic part in the
asymptotically large momentum region:
\begin{equation}
  \delta m^2(k) = - \Pi[m,k] + \Pi^{asympt}_0[m_R,k].
\end{equation}
The last symbol means the momentum dependence of the renormalized (in
any scheme) vacuum self energy for asymptotically large momenta with
mass $m_R$; so this is an explicit function of $m_R$ and $k$. If
condition \eqref{m2p} is fulfilled, this equation ensures that the
mass counterterm goes to a constant value at large momenta, since the
momentum dependence in $\delta m^2(k)$ must be slower than that of any
functions present in $\Pi^{asympt}_\MSbar[m_R,k]$.  The condition of
the constant physics reads
\begin{equation}
   m^2(k) = m^2_\MSbar + \Pi_\MSbar[m,k]-
   \Pi^{asympt}_0[m_R,k].
\end{equation}
Here again \eqref{m2p} is a self-consistent Ansatz: if it is obeyed by
$m^2(k)$, then the right hand side goes to a constant value at high
momenta faster than any functions in $\Pi^{asympt}_0[m_R,k]$.

It must be emphasized that in any of the methods above we do not have
complete 2PI resummation. To describe the high momentum regime we
still have to rely on perturbation theory; but this regime is
insensitive to matter effects and so any renormalization scheme
applicable at zero temperatures is applicable here, too. 

We still have the task to characterize the asymptotically large
momentum regime. Let us consider a two-point function in a
renormalized theory and go to this regime. Here the only dimensionful
quantity is the momentum, which is made dimensionless with help of the
renormalization scale. So we can write the self energy as
\begin{equation}
  \Sigma^\mathrm{(asym)}(k) \approx k^2 Z^\mathrm{(asym)}(k/\mu) + m^2
  Z^\mathrm{(asym)}_m(k/\mu).
\end{equation}
Therefore the dangerous terms in the asymptotic regime can be obtained
from the scale dependence of the self-energy. This means that we
should count with logarithmic dependence, and the coefficients of the
logarithms are the renormalization group beta-functions.

\section{2PI scheme in six dimensional $\Phi^3$ theory}
\label{sec:Phi3}

The simplest theory where we can demonstrate how the 2PI scheme
introduced above works, is the 6D $\Phi^3$ theory at finite
temperature. The Lagrangian of the theory is written as
\begin{equation}
  {\cal L} = \frac12 (\d\Phi)^2 -\frac{m^2}2\Phi^2 -\frac g6\Phi^3
  +\mathrm{counterterms}.
\end{equation}
Although the theory is not well defined in the path integral sense, it
behaves well in the perturbation theory, and, moreover, it shows
considerable resemblance with the perturbation theory of the gauge
models. We will use one loop level resummation; we will have momentum
dependence already at this level, so a momentum dependent resummation
is needed. On the other hand the divergence structure is rather
simple, which makes the treatment easy.

The most important peculiarity of the present treatment is that we
allow momentum dependence in the mass counterterm, which implies
momentum dependent mass. So the free Lagrangian can be written in
momentum space
\begin{equation}
  {\cal L} = \frac12 (k^2-m^2(k))\Phi(k)\Phi(-k) - \frac12 \delta
  m^2(k)\Phi(k) \Phi(-k),
\end{equation}
in such a way that the bare mass is untouched
\begin{equation}
  m^2_{bare} = m^2(k) +  \delta m^2(k) = m^2_\MSbar + \delta m^2_\MSbar.
\end{equation}

We will use real time formalism in R/A basis \cite{RAformalism}. Here
the original fields $(\Phi_1,\,\Phi_2)$ are replaced by
$(\Phi_r,\Phi_a)$ via the definition
\begin{equation}
  \Phi_1=\Phi_r +\frac12 \Phi_a,\qquad \Phi_2=\Phi_r -\frac12 \Phi_a.
\end{equation}
The propagator is a $2\times2$ matrix, where, in this basis $G_{aa}=0$
for the exact Green's function, and also $G^*_{ra}(k) = G_{ar}(k)$ is
true in the Fourier space. The interaction Lagrangian reads
\begin{equation}
  -{\cal L}_I =  \frac g6\left[\Phi_1^3- \Phi_2^3\right] = \frac g2
  \Phi_r^2 \Phi_a + \frac g{24} \Phi_a^3. 
\end{equation}

In the momentum dependent resummation we should use the most generic
possible mass matrix. Since we are in equilibrium, all the propagators
can be derived from the retarded propagator. This implies relations
between the generic self energies, which we should respect when we
choose the mass matrix. As it was pointed out in
\cite{realtimeselfenrelations}, and summarized in Appendix
\ref{sec:realtimeselfenrelations}, the self-energies satisfy
\begin{eqnarray}
  \label{sigmarels}
  && \Re \Sigma_{11} = -\Re\Sigma_{22} = \Re \Sigma_R,\qquad 
  \Im \Sigma_{11}= \Im \Sigma_{22}= (1+2n_B) \Im \Sigma_R,\nn&&
  \Sigma_{12}=-2in_B \Im \Sigma_R, \qquad 
  \Sigma_{21}=-2i(1+n_B) \Im \Sigma_R.
\end{eqnarray}
Here we used the notation $\Sigma_R=\Sigma_{ar}$ which is the retarded
self energy. This is because it parameterizes the retarded Green's
function as
\begin{equation}
  G_R^{-1}(k) = G_{R,0}^{-1}(k) - \Sigma_R(k).
\end{equation}
Choosing the mass matrix according to \eqref{sigmarels}, the only
independent parameter is the $m_R^2(k)$ retarded mass, and the free
retarded propagator reads $G_{R,0}^{-1}(k) = k^2 - m_R^2(k)$. This
implies for the free spectral function
\begin{equation}
  \label{rho0}
  \rh_0(k)  = \Disc_{k_0} G_{R,0}(k) = -2\Im  G_{R,0}(k) =  \frac{-2 \Im
    m^2_R(k)}{(k^2-\Re m^2_R(k))^2 +(\Im m^2_R(k))^2}.
\end{equation}
From the properties of $\rh_0(k)$ coming from causality, this equation
requires that $\Im m_R^2(k)$ must be an odd function of $k_0$, and
also $\Im m_R^2(k)< 0$ must be true for $k_0>0$. In the \MSbar\ scheme
it is satisfied by choosing $\Im m_R^2(k)=-\ep\, \sgn k_0$.

Since $G_{R,0}$ should be causal, $m_R^2$ must be either local or
causal, ie. Kramers-Kronig relation holds true for it:
\begin{equation}
  m_R^2(k_0,\k) = \int\limits_{-\infty}^\infty\frac{d\omega}{\pi}\frac
  {-\Im m_R^2(\omega,\k)}{k_0-\omega+i\ep} =
  \int\limits_0^\infty\frac{d\omega}{\pi}\frac {2\omega(-\Im
    m_R^2(\omega,\k))}{(k_0+i\ep)^2-\omega^2}.
\end{equation}
Therefore $\Im m_R^2$ and the local parts of $m_R^2$ will completely
determine $G_{R,0}$. 

For the counterterms we know that the infinite part is fixed, namely
\begin{equation}
  {\cal L}_{ct,div} = \frac12(k^2 \delta Z^2 -  \delta m_{div}^2)
  (\Phi_1^2-\Phi_2^2) - \frac{\delta g}6 (\Phi_1^2-\Phi_2^2)^2.
\end{equation}
This implies $\Re \delta m_R^2 = \delta m_{div}^2 +$\ finite, and $\Im
\delta m_R^2 =$\ finite. We choose for $\delta Z^2$ and $\delta g$ 
their \MSbar\ values.

If we do perturbation theory in a generic scheme, all the diagrams are
expressed through the propagators. Since all the propagators are
functionals of $\rh$, thus every diagram is a functional of $\rh$.
In particular, the one loop retarded self-energy in $\Phi^3$ model can
be written as
\begin{equation}
  \Sigma_R(k) = g^2 I_R(k)  + \delta m_R^2(k) + k^2 \delta Z^2,
\end{equation}
where $I$ is the ``bubble'' diagram
\begin{equation}
  \label{bubble}
  iI_R(k) = \pint6p iG_{rr,0}(p)\, iG_{ra,0}(k-p).
\end{equation}
At finite temperature the KMS relation implies
\begin{equation}
  iG_{rr,0}(p) = \left(\frac12 + n(p_0)\right) \rh_0(p).
\end{equation}
Thus the imaginary part of the retarded self-energy reads
\begin{equation}
  \label{imir}
  \Im I_R(k) = -\frac12 \pint6p \left(\frac12 + n(p_0)\right)
  \rh_0(p)\,\rh_0(k-p),
\end{equation}
indeed a functional of $\rh$ alone.

For later convenience, this contribution can be divided into three
parts, a ``regular'', a ``singular'' and a ``divergent'' part:
\begin{equation}
  \label{Isplit}
  I_R(k) = I_R^\mathrm{div}(k) + I_R^\mathrm{sing}(k) + I_R^\mathrm{reg}(k).
\end{equation}
The divergent piece is defined that in a certain regularization this
part goes to infinity as the regularization parameter vanishes. The
singular part is defined as a contribution that grows at large
momenta. Since the bubble contains no divergent sub-diagrams, these
contributions can be calculated from the zero temperature result. In
dimensional regularization this reads as
\begin{equation}
  I_R^{T=0}(k) =  \frac1{2(4\pi)^3} \biggl[\left(\frac1\ep - \gamma_E
    +1+\ln4\pi\right) \left(-\frac{k^2}6+m^2\right) -
  \int\limits_0^1\!dx\, (-k^2 x(1-x) + m^2) \ln\frac{-k^2 x(1-x) +
    m^2}{\mu^2}\biggr]\biggr|_{k_0\to k_0+i\ep}.
\end{equation}
In the real part we have to take the modulus of the argument of the
logarithm. The integral can be evaluated, the result reads
\begin{eqnarray}
  \label{IRreandim}
  &&  \Re I_R^{T=0}(k) = \frac1{2(4\pi)^3} \biggl[\left(\frac1\ep -
    \gamma_E +\frac83 - \ln\frac{m^2}{4\pi\mu^2} \right)
  \left(-\frac{k^2}6+m^2\right)- \frac{m^2}3
  +\frac{k^2\Gamma^3}{6}
  \ln\left|\frac{1+\Gamma}{1-\Gamma}\right|\biggr],\nn 
  && \Im I_R^{T=0}(k) =  - \Theta(k^2-4m^2)\,\frac{\sgn(k_0)}{128\pi^2}\;
  \frac{k^2\Gamma^3}6,
\end{eqnarray}
where
\begin{equation}
  \Gamma = \sqrt{1-\frac{4m^2}{k^2}}.
\end{equation}
What is particularly interesting is its large $k_0$ behavior; for a
fixed $\k$ it can be obtained as the large $k$ behavior:
\begin{eqnarray}
  &&  \Re I_R^{T=0}(k\to\infty) = \frac1{2(4\pi)^3}
  \biggl[\left(\frac1\ep - \gamma_E +\frac83 -
    \ln\frac{k^2}{4\pi\mu^2} \right) \left(-\frac{k^2}6+m^2\right)-
  \frac{m^2}2\biggr],\nn
  && \Im I_R^{T=0}(k\to\infty) = \frac{\sgn(k_0)}{128\pi^2}\;
  \left(-\frac{k^2}6 +m^2\right). 
\end{eqnarray}
For the divergent pieces we choose the following expression
\begin{equation}
  \label{Idiv}
  g^2 I_R^\mathrm{div}(k) = -k^2 \delta Z_0^2 - m^2 \delta Z_m^2,
\end{equation}
where
\begin{equation}
  \delta Z_0^2 = \frac1{12(4\pi)^3}\left(\frac1\ep - \gamma_E
    +\frac83+\ln4\pi\right),\qquad \delta Z_m^2 = -
  \frac1{2(4\pi)^3}\left(\frac1\ep - \gamma_E +\frac{13}6+\ln4\pi\right).
\end{equation}
As a singular part we choose:
\begin{equation}
  \label{Ising}
  I_R^\mathrm{sing}(k) = \frac1{2(4\pi)^3}\left(\frac{k^2}6 -
    m^2\right) \ln\frac{M^2 - k^2}{\mu^2}\biggr|_{k_0\to k_0+i\ep}.
\end{equation}
where $M^2$ is an arbitrary scale. 

The regular piece, from \eqref{Isplit}, can be obtained as
$I_R^\mathrm{reg} = I_R-I_R^\mathrm{div}-I_R^\mathrm{sing}$. Its
imaginary part, from the definition \eqref{bubble} and
from the choice of the singular part of the diagram, reads as
\begin{equation}
  \label{reg}
  \Im I_R^\mathrm{reg}(k) = -\frac12 \pint6p \left(\frac12 +
    n(p_0)\right) \rh_0(p)\,\rh_0(k-p) +
  \frac1{2(4\pi)^3}\Theta(k^2-M^2) \left(\frac{k^2}6 -
    m^2\right).
\end{equation}
From \eqref{bubble} and \eqref{imir} one can see that the real part
can be obtained from the Kramers-Kronig relation
\begin{equation}
  \Re I_R^\mathrm{reg}(k) =
  \int\limits_{-\infty}^\infty\!\frac{d\omega}{\pi} \frac{-\Im
    I_R^\mathrm{reg}(\omega,\k)}{k_0-\omega+i\ep}.
\end{equation}
This integral is convergent, since $\Im I_R^\mathrm{reg}(k)\sim
1/k_0^2$ for large $k_0$ values.

The self-energy reads with these choices as
\begin{equation}
  \Sigma_R(k) = -k^2 \delta Z_0^2 - m^2 \delta Z_m^2  + g^2
  I^\mathrm{sing}_R(k) + g^2 I^\mathrm{reg}_R(k) + \delta m_R^2(k) +
  k^2\delta Z^2.
\end{equation}

In the 2PI scheme we choose the counterterms in a specific way, namely
we choose
\begin{equation}
   \delta m_R^2(k) = m^2 \delta Z_m^2 - g^2 I^\mathrm{reg}_R(k),\qquad
   \delta Z^2 = \delta Z_0^2.
\end{equation}
Thus almost all terms are canceled in the self energy; what remains is
\begin{equation}
  \Sigma_R^{ren}(k) = g^2 I^\mathrm{sing}_R(k) =
  \frac{g^2}{2(4\pi)^3}\left(\frac{k^2}6 - m^2\right) \ln\frac{M^2 -
    k^2}{\mu^2},
\end{equation}
where the Landau prescription ($k_0\to k_0+i\ep$) is implicitly
understood in the formula. We can observe that this expression is
almost constant in the IR regime where $k^2\ll M^2$. For convenience,
in the numerical calculations we have chosen $\mu=M$, and so this
constant is $\sim \ln(1-k^2/M^2)\ll 1$ for $k^2\ll M^2$.

The full retarded propagator in this scheme reads as
\begin{equation}
  G_R^{-1}(k) = k^2 - m_R^2(k) - \frac{g^2}{2(4\pi)^3} \left(
    \frac{k^2}6 - m^2\right) \ln\frac{M^2 - k^2}{\mu^2}.
\end{equation}

Our reference scheme will be the renormalization scheme defined by the
mass counterterm $\delta m_\mathrm{ref}^2=m^2\delta Z_m^2$ and wave
function counterterm $\delta Z_\mathrm{ref}^2= \delta Z_0^2$. For
comparability of the results we have to ensure that the bare
Lagrangian is the same in the resummed and in the reference scheme.
Since the wave function renormalization is the same in the two
schemes, we have to ensure that
\begin{equation}
  Z^2 m_\mathrm{bare}^2 = m_\mathrm{ref}^2 +\delta m_\mathrm{ref}^2 =
  m_R^2(k) + \delta m_R^2(k) = m_R^2(k) +  m^2 \delta Z_m^2 - g^2
  I^\mathrm{reg}_R(k).
\end{equation}
This equation implies
\begin{equation}
  \label{gapeq}
  m_R^2(k) =  m_\mathrm{ref}^2 + g^2 I^\mathrm{reg}_R(k).
\end{equation}
This is a finite gap equation that can be solved numerically.

To determine the value of $m_\mathrm{ref}^2$ we used the
renormalization condition that the exact mass should be a predefined
value $m^2$ at zero temperature; ie. we required
\begin{equation}
  m_R^2(k^2=m^2, T=0) = m^2.
\end{equation}
From \eqref{gapeq} we can read off the value of $m_\mathrm{ref}^2$,
which later can be used at finite temperature calculations.

Before giving the results of this calculation, we mention that
formally it is also possible to choose $\mu^2=M^2-k^2$. With this
choice the 2PI scheme is exact. The validity of the above formulae can
be maintained, if we choose the same scale in the reference scheme.
Then the condition of the constant physics is modified as:
\begin{equation}
  m_R^2(k) =  m_\mathrm{ref}^2(\mu) + g^2(\mu) I^\mathrm{reg}_R(k).
\end{equation}
Here $m_\mathrm{ref}^2(\mu)$ and $g^2(\mu)$ can be explicitly
calculated using the standard renormalization group. This choice may
be advantageous in case of asymptotically free theories. In the
$\Phi^3$ model, however, we have to be careful because of the UV
Landau pole.

\subsection{Results}

For numerical calculation we used the spatial rotational invariance of
the propagators, and so at finite temperature they are two-dimensional
functions, with variables $p_0$ and $|\p|$. The integrations are
performed on 2D lattices, which are mapped with a continuous function
to the infinite 2D space.

\begin{figure}[htpb]
  \centering
  \includegraphics[height=5cm]{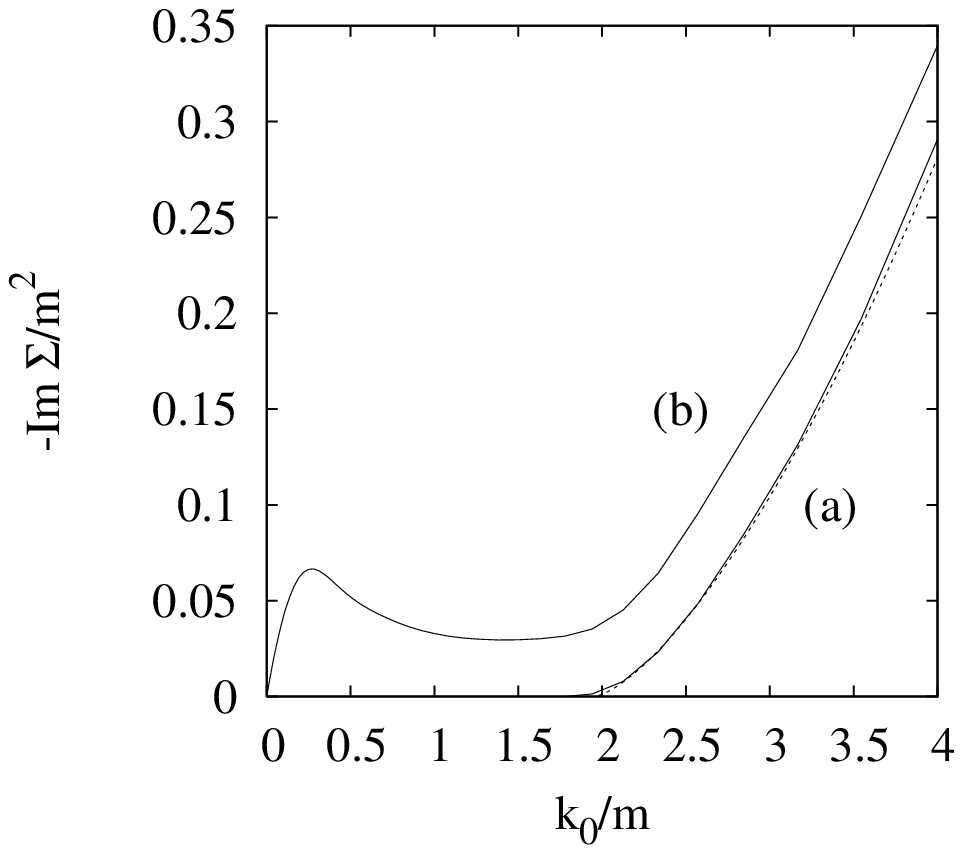}\hspace*{-1cm}
  \includegraphics[height=5cm]{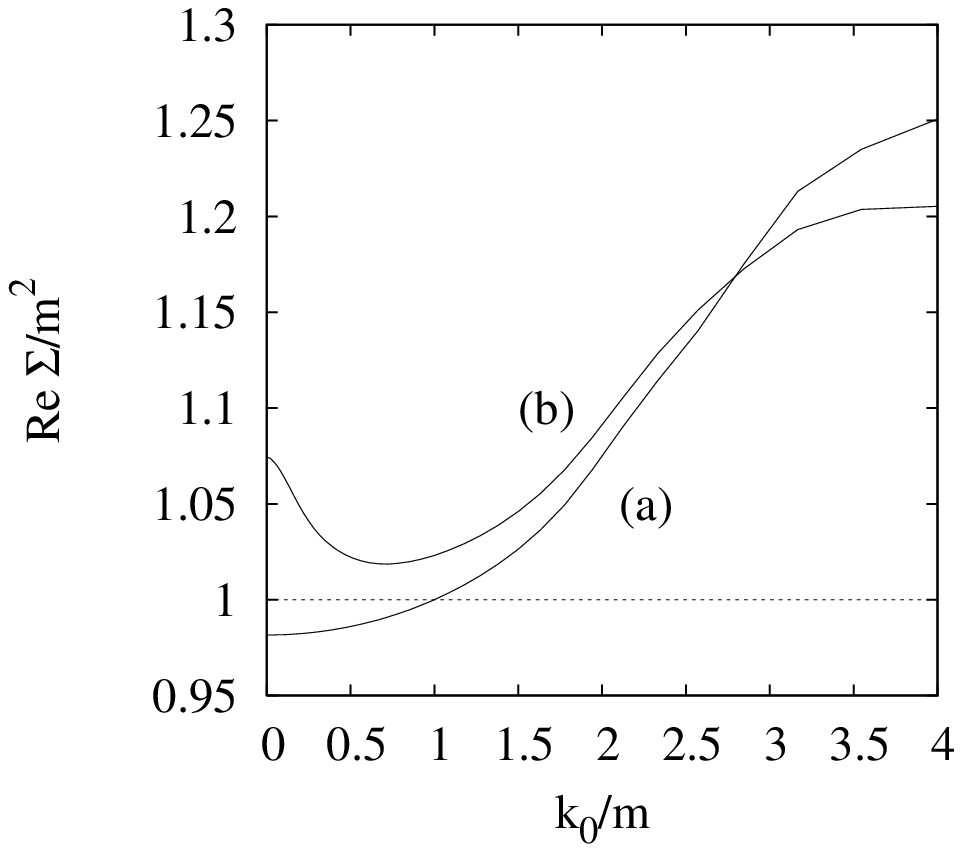}
  \caption{The imaginary and real part of the retarded self energy at zero
    temperature (a) and at finite temperature ($T/m=1$) (b). The
    spatial momentum is $k/m=0.25$ for the imaginary part case and
    $k=0$ for the real part.}
  \label{fig:PIR}
\end{figure}

The main result of the calculation is the retarded self-energy. Its
imaginary and real part at zero and at finite temperature can be seen on
 Fig. \ref{fig:PIR}.
In this run the parameters are $g=20,\;M/m=10$ and for the finite
temperature part $T/m=1$. The first plot shows the imaginary part. We
can see that at zero temperature (curve (a)) it is zero below the 2-particle
threshold $k_0/m=2$. With dotted line we plotted the one loop result
from \eqref{IRreandim}. We can see that below the 3-particle threshold
($k_0/m=3$) the agreement is rather good. The most prominent phenomena
of this plot is that at \emph{finite temperature} the spectral
function is non-vanishing everywhere (except zero) -- this is what we
expect on general arguments. One can observe the smeared-out remnant
of the Landau-damping (for $k_0<k$), but between the Landau-damping
region and the zero temperature threshold the spectral function is
still large.  Remarkably, in this regime the imaginary part is almost
constant.

On the second plot of Fig. \ref{fig:PIR} the real part of the full
self energy from the same calculation is depicted. We can see that at
zero temperature it is correctly renormalized, we indeed obtain
$\Sigma(k_0/m=1)=m^2$. At finite temperature we obtained larger real
part: this means that the temperature dependent mass correction is
positive. In the thermal mass approximation the finite temperature
curve would be above the zero temperature part by a constant shift.
But, according to this figure, there is considerable difference
between the $k_0=0$ contribution (Debye-mass) and the $k_0=m$
contribution (on-shell mass correction).

\begin{figure}[htpb]
  \centering
  \includegraphics[height=5cm]{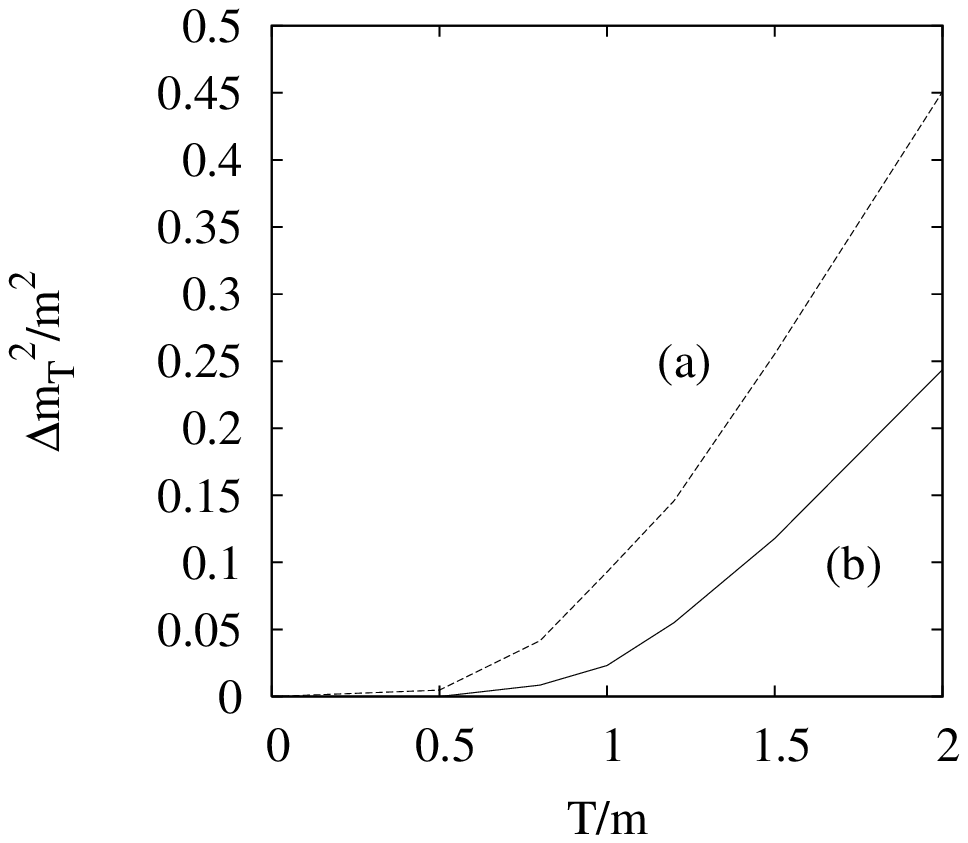}\hspace*{-1cm}
  \includegraphics[height=5cm]{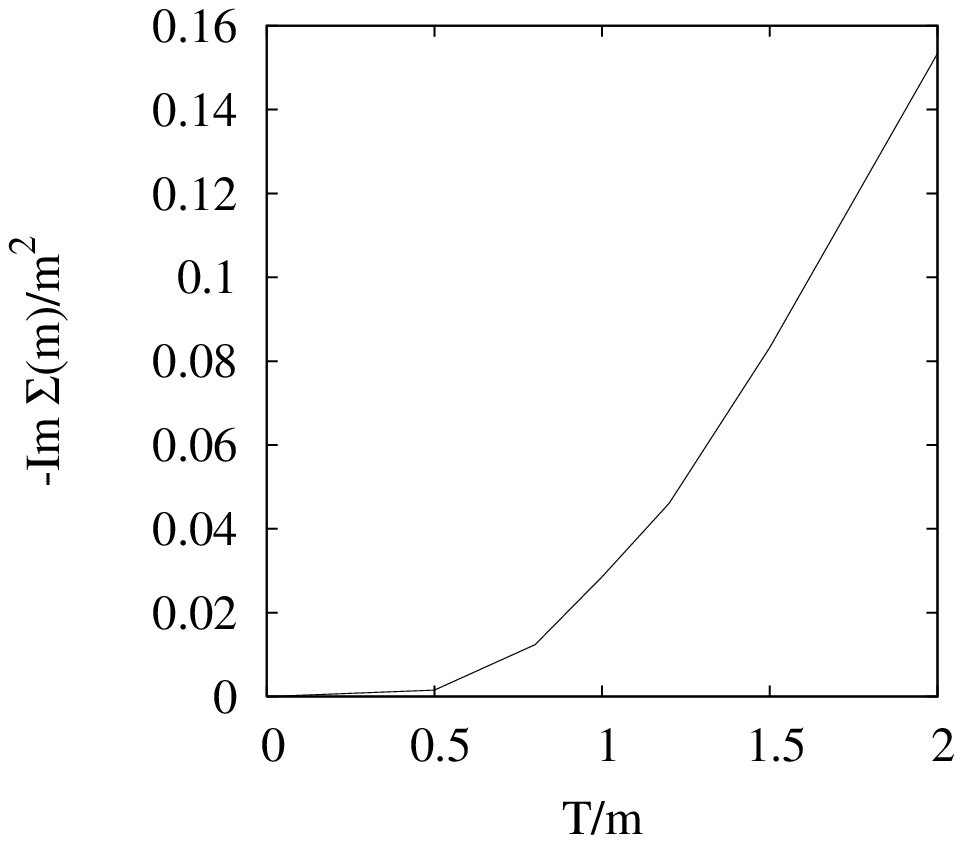}
  \caption{The first plot describes the thermal mass correction. The
    dashed curve (a) corresponds to the Debye mass (ie. $k_0=0$ case),
    the solid curve (b) is the on-shell value. In the second plot the
    on-shell imaginary part of the self energy can be seen.}
  \label{fig:PIR_T}
\end{figure}

On Fig.~\ref{fig:PIR_T} the temperature dependence of the
quasiparticle properties can be seen.
In the first plot the thermal mass correction is shown. The two curves
correspond to the thermal mass corrections evaluated at $k_0=0$ (curve
(a): this is the Debye- or screening mass) and on mass shell (curve
(b): this is the quasiparticle mass). There is a rather big difference
between the two values, which indicates that a single effective mass
term cannot be a satisfactory description. In the regime $T/m>{\cal O}(1)$
both corrections can be well fitted by a quadratic function of the
temperature ($aT^2+b$), but in the small temperature regime the curve
is much shallower. This means that although the formula
$m_\mathrm{eff}^2 \approx \bar m^2 + a T^2$ is valid at high
temperatures, but $\bar m$ is smaller than the zero temperature mass.
If this is a robust phenomenon persisting to models with spontaneous
symmetry breaking, it should lead to increasing of the critical
temperature of a phase transition as compared to the pure perturbation
theory. In the current case ($g=20$) the mass modification is about
5\%.

In the second plot of Fig.~\ref{fig:PIR_T} the on-shell value of the
imaginary part of the self energy can be seen as a function of the
temperature. This is purely resummation effect: at one loop level
there is no on-shell damping. The result again fits well to a
quadratic temperature curve ($aT^2+b$) in the regime $T/m>{\cal O}(1)$,
and it is shallower than quadratic for small temperatures. At high
temperatures it leads to a quasiparticle damping which depends
linearly on the temperature.

\section{Conclusions}
\label{sec:Conclusions}

When we leave the realm of the strict perturbative calculations and
try to implement exact, or approximate resummation formulae (like
Schwinger-Dyson equations or 2PI resummation equations), we encounter
new type of divergences in the calculation. We have then two ways to
proceed. One is that we try to find out the structure of the new type
of divergences, then resum the counterterms in the same way, thus
rendering the calculation finite. An alternative way is to realize
that divergences come solely from the ultraviolet regime which is
perturbative (if it is perturbative at zero temperature), while the
interesting results are at low energies. This allows to change the
resummation equations in a way that they do nothing at high energies,
but do resummation in the physically important IR regime.

In this alternative way the divergence structure does not vary with
the environment, which means that, in fact, we use a special
renormalization scheme. Technically this scheme works in a way that we
choose the finite parts of the counterterms (which are free to choose)
to cancel the IR part of the IR sensitive diagrams. The remaining
perturbation theory is no more IR sensitive, so we can safely use it.
The results are, however, obtained in a weird scheme which depends on
the environment, while we would need the result in a well defined,
fixed scheme (eg. in \MSbar). But results calculated in two schemes
can always be connected by renormalization group transformation: we
only have to ensure that the original bare Lagrangian is the same in
the two cases. To satisfy this requirement we have to write up
constraint equations which are, in the original resummation language,
the gap equations.

This strategy can be used for momentum dependent schemes, too. In
particular one can derive a 2PI scheme where, in the IR regime, all
the two particle reducible diagrams are canceled in the perturbation
theory. The requirement of keeping the bare Lagrangian constant can be
translated to a consistency equation for the momentum dependent mass.
If we treat the UV characteristics of the theory well, these equations
provide finite solutions.

To demonstrate in an example how this strategy works, we computed the
2PI resummed renormalized self energy in the six dimensional
$\Phi^3$ model. We used \MSbar\ scheme as reference scheme, and
we studied finite temperature dependence of different physical
quantities. The model at relatively large coupling ($g=20$) still can
be characterized by finite lifetime quasiparticles, but the spectral
function (density of states) is nonzero for all nonzero frequencies.
Interesting result is that the on-shell thermal mass and the zero
frequency (Debye) mass are quite different, in the thermal mass
correction there is almost a factor of 2 between the two values.

To pursue this project there are two ways open. Since we worked out
the technicalities, we will be able to use renormalized 2PI analysis
for other models, too. On the other hand, with the same logic we can
discuss the 4PI or even higher point irreducible approximations in a
renormalized way.

\section*{ACKNOWLEDGMENTS}
The author is grateful to A. Arrizabalaga, T.S. B\'\i r\'o, A. Patk\'os,
J. Polonyi, U. Reinosa and Zs. Sz\'ep for useful discussions. This
work was supported by Hungarian grant OTKA F043465.

\appendix

\section{The mass matrix at finite temperature}
\label{sec:realtimeselfenrelations}

Most easily we can arrive to the relations between the self-energies
in the following way. First we determine the relation between the self
energies in the CTP and R/A formalism. For this we write up the
quadratic part of the effective Lagrangian in equilibrium:
\begin{equation}
  {\cal L}_\mathrm{eff}^{(2)} = \frac12\begin{array}[c]{c}
    \left(\Phi^{(1)},\;\Phi^{(2)}\right)\cr
    \cr
    \end{array}
   \left(
    \begin{array}[c]{cc}
      k^2-\Sigma_{11}& \Sigma_{12}\cr
      \Sigma_{21} & k^2-\Sigma_{22}\cr
    \end{array}\right)  \left(\begin{array}[c]{c}
      \Phi^{(1)}\cr\Phi^{(2)}\cr \end{array}\right).
\end{equation}
To change to the R/A formalism we use $\Phi^{(1,2)} =\Phi^r
\pm\frac12\Phi^a$. Then the effective Lagrangian can be written as
\begin{equation}
   {\cal L}_\mathrm{eff}^{(2)} = \frac12\begin{array}[c]{c}
    \left(\Phi^{r},\;\Phi^{a}\right)\cr
    \cr
    \end{array}
   \left(
    \begin{array}[c]{cc}
      0& k^2-\Sigma_{ra}\cr
      k^2-\Sigma_{ar} & \Sigma_{aa}\cr
    \end{array}\right)  \left(\begin{array}[c]{c}
      \Phi^{r}\cr\Phi^{a}\cr \end{array}\right),
\end{equation}
where
\begin{eqnarray}
  && \Sigma_{11} + \Sigma_{12} + \Sigma_{21} + \Sigma_{22} = \Sigma_{rr}
  = 0\nn
  && \Sigma_{11} + \Sigma_{12} - \Sigma_{21} - \Sigma_{22} = 2\Sigma_{ar}\nn
  && \Sigma_{11} - \Sigma_{12} + \Sigma_{21} - \Sigma_{22} = 2\Sigma_{ra}\nn
  && \Sigma_{11} - \Sigma_{12} - \Sigma_{21} + \Sigma_{22} = 4\Sigma_{aa},
\end{eqnarray}
where we used the information that $G_{aa}=0$, and so
$\Sigma_{rr}=0$. The above equations can be written
\begin{eqnarray}
  &&\Sigma_R\equiv \Sigma_{ar} = \Sigma_{11} + \Sigma_{12}\nn
  &&\Sigma_A\equiv \Sigma_{ra} = \Sigma_{11} + \Sigma_{21}\nn
  &&\Sigma_K\equiv \Sigma_{aa} = \frac{\Sigma_{21} + \Sigma_{12}}2
\end{eqnarray}
On the other hand, inverting the kernel in the R/A formalism yields
the propagators
\begin{eqnarray}
  G_R = \frac1{k^2-\Sigma_R},\quad  G_A = \frac1{k^2-\Sigma_A},\quad
  iG_K= \frac{\Sigma_{aa}}{\Sigma_R-\Sigma_A}\rh,
\end{eqnarray}
where $\rh=iG_R-iG_A$. Since $G_R^*=G_A$, therefore
$\Sigma_R^*=\Sigma_A$. Moreover, since $iG_K=(\frac12+n)\rh$, so
$\Sigma_{aa}= (\frac12+n)(-2i\Im\Sigma_R)$. That means
\begin{eqnarray}
  &&\Sigma_{11}= \frac{\Sigma_R+\Sigma_A}2 -\Sigma_{aa} = \Re \Sigma_R
  +i(1+2n) \Im \Sigma_R\nn
  &&\Sigma_{22}= -\frac{\Sigma_R+\Sigma_A}2 +\Sigma_{aa} = -\Re \Sigma_R
  +i(1+2n) \Im \Sigma_R\nn
  &&\Sigma_{12} =  \frac{\Sigma_R-\Sigma_A}2 +\Sigma_{aa} =
  -2in\Im\Sigma_R \nn
  &&\Sigma_{21} = - \frac{\Sigma_R-\Sigma_A}2 +\Sigma_{aa} =
  -2i(1+n)\Im\Sigma_R.
\end{eqnarray}

\end{document}